\documentclass{sf2a-conf2013}
\usepackage{graphicx}
\usepackage{hyperref}
\usepackage[]{natbib}  
\usepackage{epstopdf}
\usepackage{MR}

\def\BibTeX{{\rm B\kern-.05em{\sc i\kern-.025em b}\kern-.08em
    T\kern-.1667em\lower.7ex\hbox{E}\kern-.125emX}}
\bibpunct{(}{)}{;}{a}{}{,}  

\begin{document}

\TitreGlobal{SF2A 2013}


\title{Present status of two-dimensional ESTER models: Application to Be
stars}

\runningtitle{ESTER models}

\author{M. Rieutord$^{1,}$}\address{Universit\'e de Toulouse; UPS-OMP; IRAP;
Toulouse, France}\address{CNRS; IRAP; 14, avenue Edouard Belin, F-31400
Toulouse, France}

\author{F. Espinosa Lara$^{1,2}$}

\setcounter{page}{1}


\maketitle


\begin{abstract}
ESTER two-dimensional models solve the steady state structure of fast rotating
early-type stars
including the large scale flows associated with the baroclinicity of the radiative
zones. Models are compared successfully to the fundamental parameters of the two
main components of the triple system $\delta$ Velorum that have been derived from
interferometric and orbit measurements. Testing the models on the Be star Achernar
($\alpha$ Eri), we cannot reproduce the data and conclude that this star has left
the main sequence and is likely crossing the Herzsprung gap. Computing main sequence
evolution of fast rotating stars at constant angular momentum shows that their
criticality increases with time suggesting that the Be phenomenon and the ensuing
mass ejections is the result of evolution.
\end{abstract}



\section{Introduction}

One of the grand challenges of stellar modelling is the design of
realistic models for fast rotating stars. Many observational data either
from interferometry, from asteroseismology, from spectroscopy (chemical
abundances) require models that are not limited to small rotation
rates. These models are necessarily two-dimensional because these stars
are no longer spherically symmetric and also because they are pervaded
by large-scale fluid flows (differential rotation and meridional
circulation) that are key elements to understand their chemical and
dynamical evolution.

Here, we wish to give a brief presentation of the first results of the
ESTER project, its promises as far as interpretation of data are
concerned and its future developments.
  
\section{The physical content of ESTER models}

Present ESTER models are describing  the internal structure of an
isolated fast rotating star in a steady state. No time evolution is
included and only early-type stars (mass larger than 1.7~\msun) can be
computed. The modeling of an outer convection zone is still a problem
that needs to be solved. The central  convective core is assumed to be
isentropic. With these prerequisites ESTER models solve the four following
partial differential equations:

\greq
     \Delta\phi = 4\pi G\rho \\
     \rho T \vv\cdot\na S = -\Div\vF + \eps_*\\
     \rho (2\vO_*\wedge\vv + \vv\cdot\na\vv) = -\na P
    -\rho\na(\phi-\textstyle{\demi}\Omega_*^2s^2)+\vF_v\\
     \Div(\rho\vv) = 0.
\egreqn{basiceq}
where we recognize the Poisson, entropy, momentum
and continuity equations respectively. They
are completed by the microphysics from OPAL and NACRE (opacities,
equation of state and reaction rates). Boundary conditions are
stress-free for the velocity field and match black body radiation for
temperature \cite[see][for details]{REL13,ELR13}.

The equations are solved using the discretization of spectral elements
in the radial direction with Chebyshev polynomial (fluid layers are of
spheroidal shape). The horizontal coordinate uses the decomposition in
spherical harmonics.  Various tests measure the quality of the numerical
solution. The ESTER code is under GNU public license and
can be downloaded at \url{http://code.google.com/p/ester-project} .

\section{Results}
 
\begin{table}
\caption{Comparison between observationally derived
parameters of the stars and tentative two-dimensional models. Data from
$\delta$ Vel are from \cite{merand_etal11}, those of Achernar are
from \cite{domiciano_etal12}. The models compare nicely with
observationally constrained data for the two components of $\delta$ Vel A
(an eclipsing binary) but have difficulties with Achernar. Here $\eps$
is the flatness, $\omega_k$ the
ratio of the equatorial angular velocity to the keplerian one, $j$ is
the mean specific angular momentum of the star and X$_\mathrm{env.}$ is
the hydrogen mass fraction in the envelope.}
\vspace*{10pt}
\centering
\begin{tabular}{lllllll}
\hline
Star                     & \multicolumn{2}{c}{Delta Velorum Aa} &
\multicolumn{2}{c}{Delta Velorum Ab} & \multicolumn{2}{c}{Achernar ($\alpha$
Eri)}\\
                      & Observations& Model  & Observations& Model &
Observations& Model\\
                         &                  &        &                  &       &
&       \\
Spectral type            & A2 IV    &      & A4 V    &       & B4 Ve  &       \\
Mass  (M$_\odot$)        & $2.43\pm0.02$ & 2.43 & 2.27$\pm0.02$ & 2.27 &            & 8.20  \\
R$_{\rm eq}$ (R$_\odot$) & 2.97$\pm$0.02 & 2.95 & 2.52$\pm$0.03 & 2.52 &
11.6$\pm$0.3   & 11.5\\
R$_{\rm pol}$ (R$_\odot$)& 2.79$\pm$0.04 & 2.77 & 2.37$\pm$0.02 & 2.36 &
8.0$\pm$0.4  & 7.9 \\
$\eps$      &         & 0.061&           & 0.064 &      & 0.310\\
$\omega_k$  &         & 0.36 &           & 0.37  &      & 0.92 \\
$i$         &     90$^\circ$ &    & 90$^\circ$&      & 101$^\circ$   &  \\
T$_{\rm eq}$ (K)         & 9450          & 9440 & 9560          & 9477 &
9955$^{+1115}_{-2339}$  & 11250\\
T$_{\rm pol}$(K)         & 10100         & 10044 & 10120        & 10115 &
18013$^{+141}_{-171}$  & 16800\\
L (L$_\odot$)            & 67$\pm$3      & 65.2 & 51$\pm$2      & 48.5 &
4500$\pm$300   & 3700\\
V$_{\rm eq}$ (km/s)      & 143           & 143  & 150           & 153 &
298$\pm$9     & 339\\
j (10$^{17}$ cm$^2$/s)    &               & 1.02 &          & 0.98 &  & 5.33 \\
k [j/R$_{\rm eq}^2\Omega_{\rm eq}$] &     & 0.0348 &      & 0.0363 & & 0.0196 \\
P$_{\rm eq}$ (days)      &               & 1.045&               & 0.832 &
& 1.72 \\
P$_{\rm pol}$ (days)     &               & 1.084&               & 0.924 &
& 1.68   \\
X$_\mathrm{env.}$        &               & 0.70 &           & 0.70 &    & 0.74\\
X$_\mathrm{core}$/X$_\mathrm{env.}$&     & 0.10 &           & 0.30 &    & 0.05\\
Z                        &               & 0.011&           & 0.011&    & 0.04\\
\hline
\end{tabular}
\label{deltavel}
\end{table}

\subsection{Modeling stars observed with optical or infra-red interferometers}

In Tab. 1 we show a comparison between ESTER models and two main
sequence stars Aa and Ab of $\delta$ Velorum. The two stars are members
of a binary system but their wide separation makes the tides of weak
influence. We first note that the model parameters nicely fit those
derived from the interferometric (and orbit) data even if for such stars
some surface convection should be expected (but it is not efficient
enough to alter the radii). These two stars are coevolving. We note
that the most massive has less hydrogen in its core as expected and that
both require the same metallicity also as expected. We note that
their mean specific angular momentum (angular momentum per unit mass) is
quite the same putting interesting constraints on the dynamics of the
formation of this system.

\subsection{Achernar and Be stars}

\begin{figure}[t!]
 \centering
 \includegraphics[width=0.8\textwidth,clip]{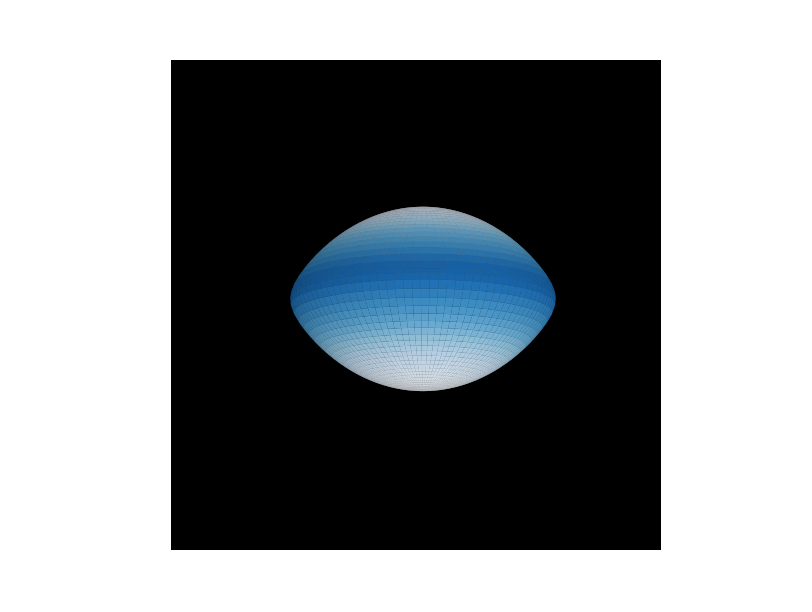}      
  \caption{A view of the surface brightness (the local flux) of
Achernar with the model of Tab.~1 (values are derived from
\citealt{domiciano_etal12}).}
  \label{achernar}
\end{figure}

The case of the Be star Achernar ($\alpha$ Eri) is quite interesting. It
has been much observed with infra-red interferometry at VLTI
\cite[e.g.][and references therein]{domiciano_etal12} since it is the
nearest Be star. As may be seen, ESTER models are not performing as well
as for $\delta$ Velorum, especially on the mass determination. Indeed, it
is known that Achernar is a binary star where the companion is an A-type
star orbiting the actual Be star in 15 yrs \cite[][]{kervella_etal08}.
Mass inferences of Achernar A is near 6.1~\msun (Domiciano de Souza, private
communication)
clearly below our value. This discrepancy may be of various origin: (i) it may well
be that Achernar has left the main sequence and is crossing
the Herzsprung gap, a configuration that is beyond reach of
ESTER models at the moment, or (ii) presently observed values are
perturbed somehow by circumstellar material including the companion. We
give in Fig.~\ref{achernar} the expected view of this star according to
the best estimate of the ESTER models.

\begin{figure}[t!]
 \centering
 \includegraphics[width=0.8\textwidth,clip]{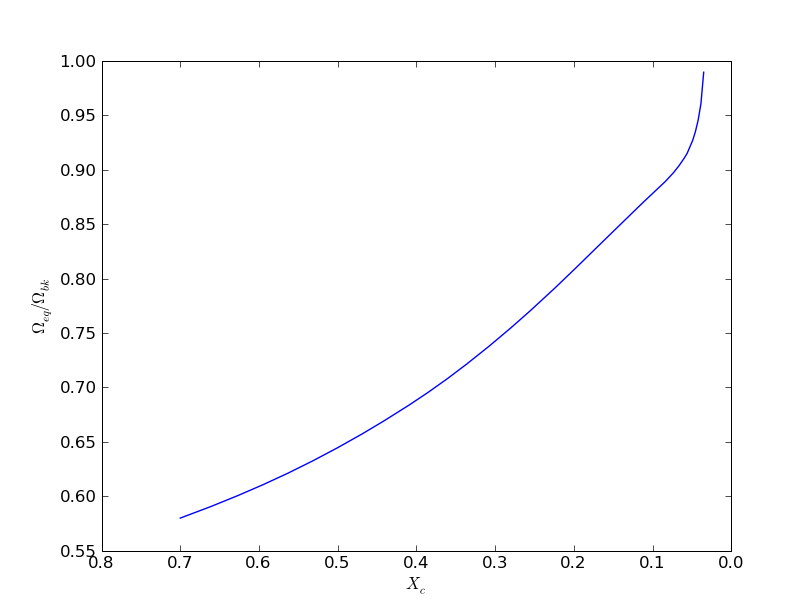}
  \caption{Equatorial angular velocity divided by the keplerian angular
velocity at equator as a function of the hydrogen mass fraction of the
star for a constant total angular momentum. Metallicity is solar and
M=7~\msun.}
  \label{OsO}
\end{figure}

ESTER models can also be used to mimic evolution of early-type stars by
decreasing the hydrogen content of the convective core. For stars that
do not lose
angular momentum, we can examine the evolution of the ratio of
angular velocity to critical angular velocity (which we call
the $\Omega$-criticality). According to
\cite{ZR12} and \cite{alecian_etal13}, A-type stars lose a
negligible amount of angular momentum and this is probably the case for
late B-type stars. In the case of zero-angular momentum loss,
Fig.~\ref{OsO} shows the evolution of $\Omega$-criticality when the star
evolves along the main sequence. We consider a 7~\msun\ model of solar
metallicity that initially rotates at $V_{\rm eq}=350$~km/s ($R_{\rm
eq}/R_{\rm pol} = 1.17$). This result suggests that Be stars result from
the evolution of initially fast rotating B stars. Mass-loss may
complicate the view, but our models show that initially fast rotating
B-stars will inevitably reach critical rotation if mass loss is weak
enough.

\section{Conclusions}

Present ESTER models can describe rather faithfully rapidly rotating
early-type main sequence stars. They therefore can be used to interpret
interferometric and asteroseismic data. In this latter case, they need
to be completed by a code that can deal with rotation non-perturbatively
(presently two codes do this job: TOP of \cite{RMJSM09} and ACOR of
\cite{ouazzani_etal12}).

In the future, developement of ESTER should include time evolution
either from chemical evolution or from gravitational contraction so as
to be able to describe early-type stars from the birthline to the giant
state. In parallel, the other challenge is to include outer convection
zone so as to be able to describe low-mass solar type stars.

\begin{acknowledgements}
The authors acknowledge the support of the French Agence Nationale de
la Recherche (ANR), under grant ESTER (ANR-09-BLAN-0140).  This work
was also supported by the Centre National de la Recherche Scientifique
(CNRS, UMR 5277), through the Programme National de Physique Stellaire
(PNPS). The numerical calculations have been carried out on the CalMip
machine of the `Centre Interuniversitaire de Calcul de Toulouse' (CICT)
which is gratefully acknowledged.
\end{acknowledgements}

\bibliographystyle{aa}  
\bibliography{/home/rieutord/tex/biblio/bibnew}

\end{document}